\documentclass[aps,twocolumn,showpacs]{revtex4}
\usepackage{graphicx}

\begin{document}
%\draft

%<<<<<<<<<<<<< TITLE >>>>>>>>>>>>>>>%
\title{Twist of  stationary black hole/ring in five dimensions}

%<<<<<<<<<<<<< AUTHOR >>>>>>>>>>>>>>>%
\author{Shinya Tomizawa$^{(1)}$, Yuki Uchida$^{(1)}$ and Tetsuya Shiromizu$^{(1,2,3)}$}

%<<<<<<<<<<<<< ADDRESS >>>>>>>>>>>>>>>%
\affiliation{$^{(1)}$Department of Physics, Tokyo Institute of Technology, 
Tokyo 152-8551, Japan}

\affiliation{$^{(2)}$Department of Physics, The University of Tokyo,  Tokyo
113-0033, Japan}

\affiliation{$^{(3)}$Advanced Research Institute for Science and Engineering, 
Waseda University, Tokyo 169-8555, Japan}

%<<<<<<<<<<<<< DATE >>>>>>>>>>>>>>>%
\date{\today}

%======================================%
%<<<<<<<<<<<<< ABSTRACT >>>>>>>>>>>>>>>%
%======================================%
\begin{abstract}
It is unlikely that uniqueness theorem holds for stationary black holes in higher dimensional spacetimes. However, 
we will examine the possibility that the higher multipole moments classify vacuum solutions uniquely. 
Especially, we compute the potentials associated with rotational Killing vectors and look at the dependence on 
the total mass $M$ and angular momentum $J$. Consequently, there is a potential $\sigma$ which 
we cannot write down in terms of integer power of $M$ and $J$ explicitly. This may be 
regarded as an evidence for the uniqueness 
using multipole moments generated by $\sigma$. 
\end{abstract}


\maketitle
%\vskip1cm

%======================================%
%<<<<<<<<<<<< SECTION I  >>>>>>>>>>>>>>%
%======================================%
%\baselineskip25pt
\label{sec:introduction}
\section{Introduction}

TeV gravity/brane world \cite{ADD} opened us the possibility of the production of higher dimensional black holes 
in accelerators \cite{LHC}. Therefore the fundamental study on higher dimensional black holes became to be 
important. Important issues are uniqueness theorem for final equilibrium state of gravitational collapse 
\cite{Israel} and no-hair theorem \cite{Price} like 
four dimensional cases. If these theorems  hold, we can use certain of exact solutions to have definite 
predictions for events in accelerators. Recently the uniqueness for non-rotating higher dimensional 
black holes in asymptotically flat spacetimes has been proven to be unique \cite{GIS,Next,Old}. See Ref. 
\cite{HSR} for a related issue of supersymmetric black holes. 
Perturbative uniqueness was also addressed in Ref. \cite{Kodama}. But, if we think of stationary cases, 
the situation is drastically changed. 
The uniqueness theorem for rotating cases does not hold due to the presence of the counterexample, 
black ring solution with $S^1 \times S^2$ event horizon, discovered by Emparan and Reall \cite{Reall} 
in five dimensions(See also Ref \cite{Roberto} for extended solutions.). As a result, there are several solutions with same total mass $M$ and angular 
momentum $J$. In stationary vacuum black hole spacetimes with two commuting rotational 
Killing vectors with $S^3$ event horizon, the uniqueness of Myers-Perry solution \cite{MP} has been proven \cite{Ida}. 

In this paper we want to discuss  the conditions for uniqueness theorem. In the theorems the asymptotic 
boundary conditions are imposed. The apparent failure in the uniqueness theorem in higher dimensions 
seems to tell us 
some missing ingredients to prove it. Here we would propose that the missing one 
is higher multipole moments. In 
four dimensional case, $M$ and $J$ are accidentally enough parameter set for the uniqueness theorem. 
If the higher order multipole moments are specified in higher dimensional spacetimes, we 
might be able to prove the uniqueness theorem. Indeed, four dimensional 
stationary spacetimes is unique under fixed multipole moments in the neighborhood 
of spatial infinity \cite{Beig}.  In higher dimensional spacetimes, on the other hand, 
the asymptotic structure is not 
so simple. In general the spacetimes will be not fixed by multipole moments in asymptotically 
flat spacetimes defined by conformal completion \cite{Spi}. But, if we focus on asymptotically flat 
spacetimes where regular Cartesian coordinate can be spanned, we can expect that spacetimes 
near spatial infinity can be fixed by multipole moments in the same way with four dimensional cases. 
Following this expectation, it is realized that the multipole moments may be able to 
distinguish black ring solutions and 
Myers-Perry solution where the regular Cartesian coordinate are spanned in asymptotic region. 

Multipole moments were firstly defined by Geroch for four dimensional static spacetimes in 
covariant way \cite{Geroch}. Then it was extended to stationary spacetimes by Hansen \cite{Hansen}. Furthermore it 
turned out that multipole moments uniquely determine the asymptotically flat and source-free 
solutions of Einstein equation in four dimensions as mention above \cite{Beig}. Geroch's multipole moments are defined in term of 
the norm of timelike Killing vector and its derivatives at spatial infinity in the framework of 
conformal completion \cite{AH}. In stationary case Hansen defined two new potential(Hansen potentials) 
which are composed of some combinations of the norm and twist potential of the timelike Killing vector. 
The moments are defined so that they satisfy a certain transformation under the 
change of conformal factor, which corresponds to the change of a origin in Newtonian limit. 
Recently Geroch's definition were extended to higher dimensional static spacetimes \cite{Spi}. 
The extension to stationary cases has not been done yet. The problem is how to find Hansen potentials 
in higher dimensions. 
As a first step for the extension to 
stationary cases, therefore, we will compute a set of potentials associated with Killing vectors. It is 
also starting point in four dimensions. 
When the exact solutions are given, yet, it will be enough for the argument of uniqueness. This can be seen as 
follows. Let write all parameters of potentials in terms of $M$ and $J$. If potentials can be expanded by some integer 
powers of $M$ and $J$ near 
spatial infinity, the solutions are degenerated. On the other hand, we cannot distinguish solutions from each other 
if potentials cannot be expanded by only integer powers of $M$ and $J$. 

The rest of this paper is organized as follows. In Sec. II, we describe 
the black string/black hole solutions and introduce the polar coordinate systems. In Sec. III, 
we define some potentials in five dimensional stationary 
vacuum spacetimes with three commuting Killing vectors. Then we will compute the scalar functions 
for five dimensional black string/black hole solutions and discuss the uniqueness properties. 
Finally we will give summary and discussion in Sec. IV.

%======================================%
%<<<<<<<<<<<< SECTION II  >>>>>>>>>>>>>>%t
%======================================%
%\baselineskip25pt
\label{sec:blackstring}
\section{Black ring and polar coordinate}

We first describe the black ring/black hole solutions \cite{Reall,MP}. 
The metric of black ring/black hole solution is written in the form \cite{Reall}
%===========<Equation>===========%
%
\begin{eqnarray}
ds^2&=&-\frac{F(x)}{F(y)}\biggl( dt+\sqrt{\frac{\nu}{\xi_1}}\frac{\xi_2-y}{A}d\Psi\biggr)^2\nonumber\\
    & & +\frac{1}{A^2(x-y)^2}\Biggl[ -F(x)\bigg(G(y)d\Psi^2+\frac{F(y)}{G(y)}dy^2 \biggr)\nonumber\\
    & & +F(y)^2\bigg(\frac{dx^2}{G(x)}+\frac{G(x)}{F(x)}d\phi^2\biggr)\Biggr], \label{metric}
\end{eqnarray}
%
%================================%
where
%===========<Equation>===========%
%
\begin{eqnarray}
& & F(\xi)=1-\frac{\xi}{\xi_1}, \\
& & G(\xi) = \nu\xi^3-\xi^2+1 \nonumber\\
& & ~~~~~~=\nu(\xi-\xi_2)(\xi-\xi_3)(\xi-\xi_4).
\end{eqnarray}
%
%================================%
The roots of $G(\xi)=0$ satisfy $\xi_2<\xi_3<\xi_4$ and $\nu \leq \nu_*=2/3{\sqrt 3}$. 
The coordinates $\Psi$ and $\phi$ are identified with period
%===========<Equation>===========%
%
\begin{eqnarray}
\Delta \phi = \Delta \Psi = \frac{4\pi {\sqrt {\xi_1-\xi_2}}}{\nu \xi^{1/2}_1(\xi_3-\xi_2)(\xi_4-\xi_2)}.
\end{eqnarray}
%
%================================%

For the black ring case, we must require 
%===========<Equation>===========%
%
\begin{eqnarray}
\xi_1 = \frac{\xi_4^2 -\xi_2 \xi_3}{2\xi_4 -\xi_2-\xi_3} \label{cond1}
\end{eqnarray}
%
%================================%
to make spacetime regular at $x=\xi_2$ and $\xi_4$. For black hole(Myers-Perry) case, 
on the other hand, 
%===========<Equation>===========%
%
\begin{eqnarray}
\xi_1 = \xi_3 \label{cond2}
\end{eqnarray}
%
%================================%
is imposed. 

Mass and angular moment are given by
%===========<Equation>===========%
%
\begin{eqnarray}
M=\frac{3\pi}{2GA^2} \frac{\xi_1 -\xi_2}{\nu \xi^2_1(\xi_3-\xi_2)(\xi_4-\xi_2)},
\end{eqnarray}
%
%================================%
%===========<Equation>===========%
%
\begin{eqnarray}
J=\frac{2\pi}{GA^3} \frac{(\xi_1-\xi_2)^{5/2}}{\nu^{3/2}\xi_1^3(\xi_3-\xi_2)^2(\xi_4-\xi_2)^2}.
\end{eqnarray}
%
%================================%
We should note that $M$ and $J$ are uniquely determined by two independent parameters $A$ and $\nu$ 
together with condition Eq. (\ref{cond1}) or (\ref{cond2}). Since $\xi_1 $ depends on the 
solutions, $M$ and $J$ are multi-valued functions of $\nu$. 

When one wants to address multipole moments, the polar coordinate is useful. 
The transformation into polar coordinate $(\rho,\chi,\theta,\mu)$ is expressed in
%===========<Equation>===========%
%
\begin{eqnarray}
& & x=\frac{\sin^2\chi\sin^2\theta}{\tilde A^2\rho^2}+\xi_2\\
& & y =-\frac{\sin^2\chi\cos^2\theta+\cos^2\chi}{\tilde A^2\rho^2}+\xi_2 \\
& & \tilde \phi = \frac{2\pi\phi}{\Delta \phi}=\mu\\
& & \tan\tilde\Psi = \frac{\cos\chi}{\sin\chi\cos\theta}
\end{eqnarray}
%
%================================%
where $\tilde \Psi :=(2\pi /\Delta \Psi) \Psi $ and 
%===========<Equation>===========%
%
\begin{eqnarray}
\tilde A:=A\frac{\xi_1\sqrt{\nu(\xi_3-\xi_2)(\xi_4-\xi_2)}}{2(\xi_1-\xi_2)}.
\end{eqnarray}
%
%================================%
The period of $\tilde \phi$ and $\tilde \Psi$ are $2\pi$. 
We can check that the metric approaches the five dimensional Minkowski spacetimes 
in polar coordinate, that is, 
%===========<Equation>===========%
%
\begin{eqnarray}
ds^2 \simeq -dt^2+d\rho^2 +\rho^2 d\Omega_3^2,
\end{eqnarray}
%
%================================%
where $ d \Omega_3^2= d \chi^2+ {\rm sin}^2 \chi (d \theta^2 + {\rm sin}^2 \theta d \mu^2)$.

%======================================%
%<<<<<<<<<<<< SECTION II  >>>>>>>>>>>>>>%
%======================================%
%\baselineskip25pt
\label{sec:appendix}
\section{Twist potential}

In this section we will define scalar functions associated with Killing vectors. The 
computation of such 
functions for exact solutions can be regarded as a first step for defining 
multipole moments. Rather say, the computation of them are enough for current purpose when 
the exact solutions are given. 

The metric of Eq. (\ref{metric}) admits three Killing vectors ;
%===========<Equation>===========%
%
\begin{eqnarray}
\xi_3^a=\biggl(\frac{\partial}{\partial\Psi }\biggr)^a, 
\xi_4^a =  \biggl(\frac{\partial}{\partial t}\biggr)^a,
\xi_5^a =\biggl(\frac{\partial}{\partial \phi}\biggr)^a.
\end{eqnarray}
%
%================================%
There are scalar potentials associated with the Killing vectors. One of them is 
%===========<Equation>===========%
%
\begin{eqnarray}
\lambda & := & -g_{ab} \xi^a_4 \xi^a_4 \nonumber \\
  & = & 1-\frac{8GM}{3\pi \rho^2} \frac{1}{1+\frac{8GM}{3\pi}\frac{{\rm sin}^2\chi {\rm cos}^2\theta +{\rm cos}^2\chi}{\rho^2} }
\end{eqnarray}
%
%================================%
Note that $\lambda$ depends on only $M$. 

To see the feature related to angular momentum, we consider twist one-form. There are 
three kinds of twist one-forms
%===========<Equation>===========%
%
\begin{eqnarray}
& & \omega_{ia}:=\epsilon_{abcde}\xi^b_4\xi^c_5\nabla^d\xi^e_i~~(i=4,5) \\
& & \sigma_{Ia}:=\epsilon_{abcde}\xi^b_3\xi^c_5\nabla^d\xi^e_I~~(I=3,5) \\
& & \tau_{\mu a}:= \epsilon_{abcde}\xi^b_4\xi^c_3 \nabla^d\xi^e_\mu~~(\mu=3,4).
\end{eqnarray}
%
%================================%
These twist one-forms are evaluated in the coordinate basis $(t,x,y,\phi, \Psi)$ as follows
%===========<Equation>===========%
%
\begin{eqnarray}
\sigma_{3a}&=&\frac{1}{A^3}\sqrt{\frac{\nu}{\xi_1}}\Biggl[\bigg\{ \frac{\nu}{\xi_1}(\xi_2-y)^2\nonumber\\
           & &- (\xi_2 -y)\biggl( \frac{F(y)G(y)}{(x-y)^2}\biggr)_{,y}-\frac{F(y)G(y)}{(x-y)^2} 
\biggr\}(dx)_a\nonumber\\
           & & +2 F(x)G(x)\frac{\xi_2-y}{(x-y)^3}(dy)_a \Biggr] \label{eq:1formsigma}
\end{eqnarray}
%
%================================%
%===========<Equation>===========%
%
\begin{eqnarray}
\omega_{4a}= -\frac{1}{A}{\sqrt {\frac{\nu}{\xi_1}}}(dx)_a
\end{eqnarray}
%
%================================%
and
%===========<Equation>===========%
%
\begin{eqnarray}
\omega_{5a} = \sigma_{5a}=\tau_{\mu a}= 0.
\end{eqnarray}
%
%================================%
By virtue of vacuum spacetime, then, there exist potentials $\sigma$ and $\omega$ 
for each twist $\sigma_{3a}$ and $ \omega_{4a}$ as 
 %===========<Equation>===========%
%
\begin{eqnarray}
\sigma_{3 a}=\nabla_a\sigma~~{\rm and}~~\omega_{4 a}=\nabla_a\omega. 
\end{eqnarray}
%
%================================%
We obtain these twist potentials from the above twist one-forms by integrating them. The result are 
%===========<Equation>===========%
%
\begin{eqnarray}
\sigma&=&\frac{1}{\xi_1A^3}\sqrt{\frac{\nu}{\xi_1}}\Biggr[\frac{(\xi_1-x)(x-2y+\xi_2)(\nu x^3-x^2+1)}{(x-y)^2}\nonumber\\
      & & ~~~~~~~~~~+\nu x^3-(\nu\xi_1-\nu\xi_2+1)x^2 \nonumber \\
      & & ~~~~~~~~~~-(-\xi_1+\xi_2+\nu\xi_1\xi_2-\nu\xi_2^2)x\Biggl]
\end{eqnarray}
%
%================================%
and
%===========<Equation>===========%
%
\begin{eqnarray}
\omega= -\frac{1}{A}{\sqrt {\frac{\nu}{\xi_1}}}(x-\xi_2).
\end{eqnarray}
%
%================================%
Here it is reminded that the period of coordinates $\phi$ and $\Psi$ are $\Delta \phi = \Delta \Psi \neq 2\pi$. 
Thus it is better to rewire down $\sigma$ and $\omega$ 
in new coordinates $\tilde \phi $ and $\tilde \Psi$ and the rescaled potentials are 
used. In polar coordinate they become 
%===========<Equation>===========%
%
\begin{eqnarray}
\tilde \sigma &=& \frac{\partial \phi}{\partial \tilde \phi} \frac{\partial \Psi}{\partial \tilde \Psi}\sigma \nonumber \\
              &=& \frac{\Delta \phi}{2\pi} \frac{\Delta \Psi}{2\pi} \sigma \nonumber\\
              &=&\frac{16G^2MJ}{3\pi^2\rho^2}\sqrt{\frac{\xi_1}{\xi_1-\xi_2}}\sin^2\chi\sin^2\theta\bigg\{\sin^2\chi\sin^2\theta (1\nonumber\\
              & &+\sin^2\chi\cos^2\theta+\cos^2\chi)-1\biggl\}\Biggl[ -\xi_1+3\xi_2-6\nu\xi_2^2\nonumber\\
              & &-3\nu\xi_1\xi_2+\frac{8GM}{3\pi\rho^2}(\xi_1-\xi_2)(1+\nu\xi_1-4\nu\xi_2)\nonumber\\
              & &\times\sin^2\chi\sin^2\theta-\biggl( \frac{8GM}{3\pi\rho^2}(\xi_1-\xi_2)\biggr)^2\nu\nonumber\\
              &&\times\sin^4\chi\sin^4\theta\Biggr] \label{twistpot}
\end{eqnarray}
%
%================================%
and
%===========<Equation>===========%
%
\begin{eqnarray}
\tilde \omega = \frac{\partial \phi}{\partial \tilde \phi} \omega 
= \frac{\Delta \phi}{2\pi} \omega = -\frac{4GJ}{\pi} \frac{{\rm sin}^2\chi {\rm sin}^2 \theta}{\rho^2}.
\end{eqnarray}
%
%================================%
%===========<Equation>===========%
%
%\begin{eqnarray}
%\tilde \sigma &=& \frac{\Delta \phi}{2\pi} \frac{\Delta \Psi}{2\pi} \sigma \nonumber\\
%              &=&\frac{1028A^3\nu^{3/2}\xi_1^{7/2}G^3M^6}{729\pi^4J^2}\biggl(\frac{27\pi J^2}{8A^2G\ \xi_1^2\nu M^3\rho^2}\biggr)\nonumber\\
%              & & \times\Biggl[ \bigg\{2\sin^4\chi\sin^4\theta (1+\sin^2\chi\cos^2\theta+\cos^2\chi)(-\xi_1\nonumber\\
%              & &+3\xi_2-3\nu\xi_2^2+3\nu\xi_1\xi_2-3\nu\xi_2)+\sin\chi^2\sin^2\theta (-\xi_1\nonumber\\
%              & &+\xi_2-2\nu\xi_1+2\xi_2-\xi_2^2+3\nu\xi_2^2+\nu\xi_1\xi_2-2) \bigg\}\nonumber\\
%              & &+\frac{27\pi J^2}{8A^2G\ \xi_1^2\nu M^3\rho^2}\biggl\{ 2\sin^6\chi\sin^6\theta (1+\sin^2\chi\cos^2\theta\nonumber\\
%              & &+\cos^2\chi)(1+\nu\xi_1-7\nu\xi_2)+\sin^4\chi\sin^4\theta (-\nu\xi_1\nonumber\\
%              & &+4\nu\xi_2-1)\biggr\}+\biggl( \frac{27\pi J^2}{8A^2G\ \xi_1^2\nu M^3\rho^2}\biggr)^2\nonumber\\
%              & &\times\bigg\{-2\sin^8\chi\sin^8\theta (1+\sin^2\chi\cos^2\theta+\cos^2\chi)\nonumber \\
%              & &+\sin^6\chi\sin^6\theta  \biggr\}\nu \Biggr]. \label{twistpot}
%\end{eqnarray}
%
%================================%

Let us consider the situation with fixed $M$ and $J$. The functions $\lambda$ and $\tilde \omega$ defined above are 
written in terms of $M$ and $J$. Moreover $M$ and $J$ are contained as the form of integer powers of them in 
asymptotic region. Therefore we cannot use $\lambda$ and $\tilde \omega$ to classify the 
solutions with same $M$ and $J$. 
On the other hand, it is easy to see from Eq. (\ref{twistpot}) that 
the dependence of $\sigma$ on $M$ and $J$ are quite non-trivial. 
If we consider a certain exact solution, $\sigma$ can be written in terms of $M$ and $J$. However, 
$M$ and $J$ are not included as the form of integer powers in asymptotic region. Since the dependence depends on solutions, 
$\sigma$ are not fixed even if we fixed $M$ and $J$. Therefore we can split the degeneracy between 
black holes and black rings by twist potential $\sigma$. To confirm our this argument, 
we should perform numerical evaluation the value of coefficients, which cannot be explicitly written in terms of $M$ and $J$ 
in Eq. (\ref{twistpot}), for each solutions. For example, pick up the coefficient of the term of $\rho^{-6}$, 
${\sqrt {\frac{\xi_1}{\xi_1-\xi_2}}}(\xi_1-\xi_2)^2 \nu$. The result is in Fig. 1. 
From this result we conclude the twist potential $\sigma$ has different profiles depending on each 
solutions for fixed mass $M$ and angular momentum $J$.

\begin{figure}[htbp]
\begin{center}
\includegraphics[width=1.0\linewidth]{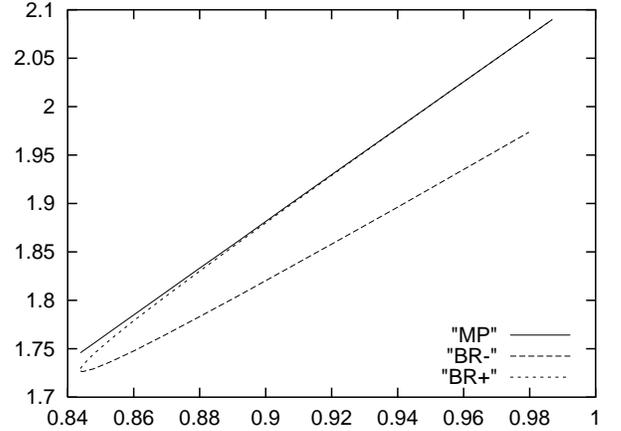}
\end{center}
\caption{The above numerical result shows the dependence on solutions of 
the twist potential $\sigma$. Here, we computed the coefficient of the 
term of $\rho^{-6}$ in the twist potential $\sigma$.  We denote it as 
$\alpha=\xi_1^{\frac{1}{2}}(\xi_1-\xi_2)^{\frac{3}{2}}\nu $.  
The vertical axis is $\alpha$ and the horizontal axis is $(27\pi /32G)J^2/M^3$. The 
dotted line and the dashed line correspond to the two black 
ring solutions ( These solutions are different in the values of $\nu$ [$\nu({\rm BR}+)>\nu({\rm BR}-)$].). 
The solid line expresses Myers-Perry solution. This result also shows that 
the twist potential of Myers-Perry solutions is asymptotically close to 
the one of the black ring solution as $(27\pi /32G)J^2/M^3 \to 1$, where the black ring and Myers-Perry black hole degenerate\cite{Reall}. 
 } 
\end{figure}

\section{conclusion}

In this paper we evaluated a twist potential $\sigma$ for stationary five dimensional black ring/black hole. 
As a result we saw that its shape depends on solutions. Therefore this result means that using twist 
potentials we can distinguish 
the black ring solution from black hole one with same mass and angular momentum. Yet, it indicates that 
a sort of uniqueness theorem may hold under the strong asymptotic conditions specified by 
the multipole moments defined via this $\sigma$. 
 
As future work, the definition of the multipole moments in higher dimension will be important. 
It is expected that they uniquely 
determine stationary solutions. It is a open question how Hansen's 
multipole moments are constructed from the combinations of twist potentials and gravitational potentials. 
We also should consider the multipole moments in  Einstein-Maxwell(or higher form fields) systems in four  or higher dimensional 
spacetimes because black holes produced in accelerator have charges in general.

\section*{Acknowledgements}

The works of ST and YU were supported 
by the 21st Century COE Program at TokyoTech "Nanometer-Scale Quantum Physics" supported 
by the Ministry of Education, Culture, Sports, Science and Technology. 
The work of TS was supported by Grant-in-Aid for Scientific 
Research from Ministry of Education, Science, Sports and Culture of 
Japan(No.13135208, No.14740155 and No.14102004). 

%\vskip 1cm 

\appendix

\section{calculation}

In this Appendix we write down useful formulae and sketch of the derivation of twist potential $\sigma$ of 
Eq. (\ref{twistpot}). 

\subsection{Inverse and determinant of metric}
The inverse of metric of Eq. (\ref{metric}) is 
%===========<Equation>===========%
%
\begin{eqnarray}
g^{ab}&=&-\Bigg[\frac{\nu (x-y)^2(\xi_2-y)^2}{\xi_1 F(x)G(y)}+\frac{F(y)}{F(x)}\Biggr]\biggl( \frac{\partial}{\partial t}\biggr)^a\biggl( \frac{\partial }{\partial t}\biggr)^b\nonumber\\
      & &+2A\sqrt{\frac{\nu}{\xi_1}}\frac{(x-y)^2(\xi_2-y)}{F(x)G(y)}\biggl( \frac{\partial}{\partial t}\biggr)^{(a}\biggl( \frac{\partial }{\partial \Psi}\biggr)^{b)}\nonumber\\
      & &-\frac{A^2(x-y)^2}{F(x)G(y)}\biggl( \frac{\partial}{\partial \Psi}\biggr)^a\biggl( \frac{\partial }{\partial \Psi}\biggr)^b\nonumber\\
      & &+\frac{A^2(x-y)^2G(x)}{F(y)^2}\biggl( \frac{\partial}{\partial x}\biggr)^a\biggl( \frac{\partial }{\partial x}\biggr)^b\nonumber\\
      & &-\frac{A^2(x-y)^2G(y)}{F(x)F(y)}\biggl( \frac{\partial}{\partial y}\biggr)^a\biggl( \frac{\partial }{\partial y}\biggr)^b\nonumber\\
      & &\frac{A^2(x-y)^2F(x)}{F(y)^2G(x)}\biggl( \frac{\partial}{\partial \phi}\biggr)^a\biggl( \frac{\partial }{\partial \phi}\biggr)^b.
\end{eqnarray}
%
%================================%
The determinant $g={\rm det}\ g_{\mu\nu}$ of the metric $g_{\mu\nu}$ is
%===========<Equation>===========%
%
\begin{eqnarray}
g=-\frac{F(x)^2F(y)^4}{A^8(x-y)^8}.
\end{eqnarray}
%
%================================%

\subsection{Twist potential}

In this subsection, we sketch the calculation of twist potential $\sigma$. From Eq.(\ref{eq:1formsigma}), the partial differential equations for $\sigma$ are
%===========<Equation>===========%
%
\begin{eqnarray}
\sigma_{,x}&=&\frac{1}{A^3}\sqrt{\frac{\nu}{\xi_1}}\bigg\{ \frac{\nu}{\xi_1}(\xi_2-y)^2\nonumber\\
           & &- (\xi_2 -y)\biggl( \frac{F(y)G(y)}{(x-y)^2}\biggr)_{,y}-\frac{F(y)G(y)}{(x-y)^2}, \label{eq:sigmax}
\biggr\}
\end{eqnarray}
%
%================================%
and
%===========<Equation>===========%
%
\begin{eqnarray}
\sigma_{,y}=\frac{2}{A^3}\sqrt{\frac{\nu}{\xi_1}} F(x)G(x)\frac{\xi_2-y}{(x-y)^3}\label{eq:sigmay}.
\end{eqnarray}
%
%================================%
Integrating Eq. (\ref{eq:sigmay}), we obtain 
%===========<Equation>===========%
%
\begin{eqnarray}
\sigma &=&\frac{1}{A^3\xi_1}\sqrt{\frac{\nu}{\xi_1}}\Biggl[(\xi_1-x)(\nu x^3-x^2+1)\Biggl\{\frac{1}{(x-y)}\nonumber\\
       & &+\frac{\xi_2-y}{(x-y)^2} \Biggr\}+f(x)\Biggr], \label{eq:sigma}
\end{eqnarray}
%
%================================%
where $f(x)$ is an arbitrary function of $x$. 
Substituting Eq. (\ref{eq:sigma}) into Eq. (\ref{eq:sigmax}), we obtain the ordinal differential equation 
for $f(x)$ as 
%===========<Equation>===========%
%
\begin{eqnarray}
f'(x)&=&3\nu x^2+(2\nu\xi_2-2\nu\xi_1-2)x\nonumber\\
     & &+\xi_1-\xi_2-\nu\xi_1\xi_2+\nu\xi_2^2.
\end{eqnarray}
%
%================================%
This can be easily integrated and then 
%===========<Equation>===========%
%
\begin{eqnarray}
f(x)&=&\nu x^3+(\nu\xi_2-\nu\xi_1-1)x^2\nonumber\\
    & &+(\xi_1-\xi_2-\nu\xi_1\xi_2+\nu\xi_2^2)x+{\rm const}. \label{solfx}
\end{eqnarray}
%
%================================%
Finally we obtain Eq. (\ref{twistpot}) substituting Eq. (\ref{solfx}) into (\ref{eq:sigma}).

\end{document}